\documentstyle{aipproc}
\begin{document}
%\draft

\def\simge{\hspace*{0.2em}\raisebox{0.5ex}{$>$}
     \hspace{-0.8em}\raisebox{-0.3em}{$\sim$}\hspace*{0.2em}}
\def\simle{\hspace*{0.2em}\raisebox{0.5ex}{$<$}
     \hspace{-0.8em}\raisebox{-0.3em}{$\sim$}\hspace*{0.2em}}
\def\bra#1{{\langle#1\vert}}
\def\ket#1{{\vert#1\rangle}}
\def\coeff#1#2{{\scriptstyle{#1\over #2}}}
\def\undertext#1{{$\underline{\hbox{#1}}$}}
\def\hcal#1{{\hbox{\cal #1}}}
\def\sst#1{{\scriptscriptstyle #1}}
\def\eexp#1{{\hbox{e}^{#1}}}
\def\rbra#1{{\langle #1 \vert\!\vert}}
\def\rket#1{{\vert\!\vert #1\rangle}}
\def\lsim{{ <\atop\sim}}
\def\gsim{{ >\atop\sim}}
\def\nubar{{\bar\nu}}
\def\psibar{{\bar\psi}}
\def\Gmu{{G_\mu}}
\def\alr{{A_\sst{LR}}}
\def\wpv{{W^\sst{PV}}}
\def\evec{{\vec e}}
\def\notq{{\not\! q}}
\def\notl{{\not\! \ell}}
\def\notk{{\not\! k}}
\def\notp{{\not\! p}}
\def\notpp{{\not\! p'}}
\def\notder{{\not\! \partial}}
\def\notcder{{\not\!\! D}}
\def\notA{{\not\!\! A}}
\def\notv{{\not\!\! v}}
\def\Jem{{J_\mu^{em}}}
\def\Jana{{J_{\mu 5}^{anapole}}}
\def\nue{{\nu_e}}
\def\mn{{m_\sst{N}}}
\def\mns{{m^2_\sst{N}}}
\def\me{{m_e}}
\def\mes{{m^2_e}}
\def\mq{{m_q}}
\def\mqs{{m_q^2}}
\def\mz{{M_\sst{Z}}}
\def\mzs{{M^2_\sst{Z}}}
\def\mws{{M^2_\sst{W}}}
\def\ubar{{\bar u}}
\def\dbar{{\bar d}}
\def\sbar{{\bar s}}
\def\qbar{{\bar q}}
\def\sstw{{\sin^2\theta_\sst{W}}}
\def\gv{{g_\sst{V}}}
\def\ga{{g_\sst{A}}}
\def\pv{{\vec p}}
\def\pvs{{{\vec p}^{\>2}}}
\def\ppv{{{\vec p}^{\>\prime}}}
\def\ppvs{{{\vec p}^{\>\prime\>2}}}
\def\qv{{\vec q}}
\def\qvs{{{\vec q}^{\>2}}}
\def\xv{{\vec x}}
\def\xpv{{{\vec x}^{\>\prime}}}
\def\yv{{\vec y}}
\def\tauv{{\vec\tau}}
\def\sigv{{\vec\sigma}}
\def\sst#1{{\scriptscriptstyle #1}}
\def\gpnn{{g_{\sst{NN}\pi}}}
\def\grnn{{g_{\sst{NN}\rho}}}
\def\gnnm{{g_\sst{NNM}}}
\def\hnnm{{h_\sst{NNM}}}

\def\xivz{{\xi_\sst{V}^{(0)}}}
\def\xivt{{\xi_\sst{V}^{(3)}}}
\def\xive{{\xi_\sst{V}^{(8)}}}
\def\xiaz{{\xi_\sst{A}^{(0)}}}
\def\xiat{{\xi_\sst{A}^{(3)}}}
\def\xiae{{\xi_\sst{A}^{(8)}}}
\def\xivtez{{\xi_\sst{V}^{T=0}}}
\def\xivteo{{\xi_\sst{V}^{T=1}}}
\def\xiatez{{\xi_\sst{A}^{T=0}}}
\def\xiateo{{\xi_\sst{A}^{T=1}}}
\def\xiva{{\xi_\sst{V,A}}}

\def\rvz{{R_\sst{V}^{(0)}}}
\def\rvt{{R_\sst{V}^{(3)}}}
\def\rve{{R_\sst{V}^{(8)}}}
\def\raz{{R_\sst{A}^{(0)}}}
\def\rat{{R_\sst{A}^{(3)}}}
\def\rae{{R_\sst{A}^{(8)}}}
\def\rvtez{{R_\sst{V}^{T=0}}}
\def\rvteo{{R_\sst{V}^{T=1}}}
\def\ratez{{R_\sst{A}^{T=0}}}
\def\rateo{{R_\sst{A}^{T=1}}}

\def\mro{{m_\rho}}
\def\mks{{m_\sst{K}^2}}
\def\mpi{{m_\pi}}
\def\mpis{{m_\pi^2}}
\def\mom{{m_\omega}}
\def\mphi{{m_\phi}}
\def\Qhat{{\hat Q}}

\def\FOS{{F_1^{(s)}}}
\def\FTS{{F_2^{(s)}}}
\def\GAS{{G_\sst{A}^{(s)}}}
\def\GES{{G_\sst{E}^{(s)}}}
\def\GMS{{G_\sst{M}^{(s)}}}
\def\GATEZ{{G_\sst{A}^{\sst{T}=0}}}
\def\GATEO{{G_\sst{A}^{\sst{T}=1}}}
\def\mdax{{M_\sst{A}}}
\def\mustr{{\mu_s}}
\def\rsstr{{r^2_s}}
\def\rhostr{{\rho_s}}
\def\GEG{{G_\sst{E}^\gamma}}
\def\GEZ{{G_\sst{E}^\sst{Z}}}
\def\GMG{{G_\sst{M}^\gamma}}
\def\GMZ{{G_\sst{M}^\sst{Z}}}
\def\GEn{{G_\sst{E}^n}}
\def\GEp{{G_\sst{E}^p}}
\def\GMn{{G_\sst{M}^n}}
\def\GMp{{G_\sst{M}^p}}
\def\GAp{{G_\sst{A}^p}}
\def\GAn{{G_\sst{A}^n}}
\def\GA{{G_\sst{A}}}
\def\GETEZ{{G_\sst{E}^{\sst{T}=0}}}
\def\GETEO{{G_\sst{E}^{\sst{T}=1}}}
\def\GMTEZ{{G_\sst{M}^{\sst{T}=0}}}
\def\GMTEO{{G_\sst{M}^{\sst{T}=1}}}
\def\lamd{{\lambda_\sst{D}^\sst{V}}}
\def\lamn{{\lambda_n}}
\def\lams{{\lambda_\sst{E}^{(s)}}}
\def\bvz{{\beta_\sst{V}^0}}
\def\bvo{{\beta_\sst{V}^1}}
\def\Gdip{{G_\sst{D}^\sst{V}}}
\def\GdipA{{G_\sst{D}^\sst{A}}}
\def\fks{{F_\sst{K}^{(s)}}}
\def\FIS{{F_i^{(s)}}}
\def\fpi{{F_\pi}}
\def\fk{{F_\sst{K}}}

\def\RAp{{R_\sst{A}^p}}
\def\RAn{{R_\sst{A}^n}}
\def\RVp{{R_\sst{V}^p}}
\def\RVn{{R_\sst{V}^n}}
\def\rva{{R_\sst{V,A}}}
\def\xbb{{x_B}}

\def\mlq{{M_\sst{LQ}}}
\def\mlqs{{M_\sst{LQ}^2}}
\def\lscal{{\lambda_\sst{S}}}
\def\lvect{{\lambda_\sst{V}}}

\def\PR#1{{{\em   Phys. Rev.} {\bf #1} }}
\def\PRC#1{{{\em   Phys. Rev.} {\bf C#1} }}
\def\PRD#1{{{\em   Phys. Rev.} {\bf D#1} }}
\def\PRL#1{{{\em   Phys. Rev. Lett.} {\bf #1} }}
\def\NPA#1{{{\em   Nucl. Phys.} {\bf A#1} }}
\def\NPB#1{{{\em   Nucl. Phys.} {\bf B#1} }}
\def\AoP#1{{{\em   Ann. of Phys.} {\bf #1} }}
\def\PRp#1{{{\em   Phys. Reports} {\bf #1} }}
\def\PLB#1{{{\em   Phys. Lett.} {\bf B#1} }}
\def\ZPA#1{{{\em   Z. f\"ur Phys.} {\bf A#1} }}
\def\ZPC#1{{{\em   Z. f\"ur Phys.} {\bf C#1} }}
\def\etal{{{\em   et al.}}}

\def\delalr{{{delta\alr\over\alr}}}
\def\pbar{{\bar{p}}}
\def\lamchi{{\Lambda_\chi}}

\def\qw0{{Q_\sst{W}^0}}
\def\qwp{{Q_\sst{W}^P}}
\def\qwn{{Q_\sst{W}^N}}
\def\qwe{{Q_\sst{W}^e}}

\def\gae{{g_\sst{A}^e}}
\def\gve{{g_\sst{V}^e}}
\def\gvf{{g_\sst{V}^f}}
\def\gaf{{g_\sst{A}^f}}
\def\gvu{{g_\sst{V}^u}}
\def\gau{{g_\sst{A}^u}}
\def\gvd{{g_\sst{V}^d}}
\def\gad{{g_\sst{A}^d}}

\def\gvftil{{\tilde g_\sst{V}^f}}
\def\gaftil{{\tilde g_\sst{A}^f}}
\def\gvetil{{\tilde g_\sst{V}^e}}
\def\gaetil{{\tilde g_\sst{A}^e}}
\def\gvqtil{{\tilde g_\sst{V}^e}}
\def\gaqtil{{\tilde g_\sst{A}^e}}
\def\gvutil{{\tilde g_\sst{V}^e}}
\def\gautil{{\tilde g_\sst{A}^e}}
\def\gvdtil{{\tilde g_\sst{V}^e}}
\def\gadtil{{\tilde g_\sst{A}^e}}

\def\hvf{{h_\sst{V}^f}}
\def\hvu{{h_\sst{V}^u}}
\def\hvd{{h_\sst{V}^d}}
\def\hve{{h_\sst{V}^e}}
\def\hvq{{h_\sst{V}^q}}

\def\delp{{\delta_P}}
\def\delzp{{\delta_{00}}}
\def\deld{{\delta_\Delta}}
\def\dele{{\delta_e}}

\def\dnewmu{{\Delta_\mu^\sst{NEW}}}
\def\dnewbeta{{\Delta_\beta^\sst{NEW}}}
\def\dnewpv{{\Delta_\sst{PV}^\sst{NEW}}}

\def\apv{{A_\sst{PV}}}
\def\apvnsid{{A_\sst{PV}^\sst{NSID}}}
\def\qpv{{Q_\sst{W}}}

\title{
Electrons, New Physics, and the Future of Parity-Violation}

\author{M.J. Ramsey-Musolf$^{a,b}$
%\thanks{National Science Foundation
%Young Investigator}
\\[0.3cm]
}
\address{
$^a$ Department of Physics, University of Connecticut,
Storrs, CT 06269 USA\\
$^b$ Theory Group, Thomas Jefferson National Laboratory, Newport News,
VA 23606 USA
}

%\date

\maketitle

\begin{abstract}
The study of parity-violation in semi-leptonic processes has yielded important
insights into the structure of the Standard Model and the substructure of the
nucleon. I discuss the future of semi-leptonic parity-violation and the role it
might play in uncovering physics beyond the Standard Model.

\end{abstract}

%\pacs{11.30.Er, 12.15.Ji, 25.30.Bf}

%\narrowtext

\section{Introduction}
\label{sec:intro}

In addition to celebrating the silver anniversary this year of electron
scattering
at the MIT-Bates Laboratory, we may also mark the passing of 25 years for
another
sub-field of physics: parity-violation (PV) in semi-leptonic neutral current
interactions. Since the MIT-Bates Laboratory has made important
contributions to
the field of neutral current PV, it seems highly appropriate to consider
the future
of the field at this Symposium. Paul Souder has discussed in detail the
history of neutral current PV in electron scattering, and Besty Beise has
summarized the present program of strange-quark searches here at MIT-Bates, the
Jefferson Lab, and Mainz. Consequently, I will focus on the future: where
the field
might go once the current round of parity-violating electron scattering (PVES)
experiments are completed. I will also broaden the topic to include PV in
atoms.
Historically, atomic parity-violation (APV) has been at the forefront of
the field,
and it will undoubtedly continue to hold such a position in the future. In
discussing the future, I hope to convey the following three points: (a) the
forefront of neutral current PV will consist of searches for physics beyond the
Standard Model; (b) APV and PVES can play complementary roles in this
search for
\lq\lq new physics"; and (c) parity-violating, low-energy semi-leptonic
processesand high-energy collider searches can, in principle, provide complementary 
insights
as to what may lie beyond the Standard Model (SM). My guess is that this
situation
will persist for the better part of the next decade, until the LHC begins
to produce
significant physics results.

Before considering the next decade, it is useful to look back briefly a the
last
quarter century. One may trace the birth of this field to the Bouchiats, who
proposed in 1974 that studying PV atomic processes might produce evidence
for the
weak neutral currents of the SM in the semi-leptonic sector \cite{Bou74}. The
Bouchiats suggested a clever technique for enhancing the signal for these tiny
neutral currents so that they might be observed in table top experiments. This
technique, called \lq\lq Stark mixing", relies on the interference of a
Stark-induced mixing of opposite parity states in an atom and the mixing
caused by
weak neutral currents. In effect, the Stark-induced amplitude functions as
a lever
arm to magnify the importance of the neutral current amplitude. The
importance of
this idea cannot be overstated. Following the Bouchiats' proposal, a number of
groups endeavored to search for weak neutral currents in APV, using either the
Stark-mixing idea or by studying the rotation of plane-polarized light as
it passes
through a gas of atoms (see, {\em e.g.}, Ref. \cite{Bud98} and references
therein).
In fact, the recent, very precise result for cesium APV reported by the Boulder
group was obtained using a variation on the Bouchiats' original
Stark-mixing idea
\cite{Woo97}. The result of these APV experiments has been to confirm the SM
prediction for the structure of the weak neutral current in the low-energy
domain
at the few percent level. Given the scope of effort involved in testing the
SM in
high-energy collider experiments, the results of the APV measurements
represent a
significant triumph for table top physics.

Among noteworthy collider experiments are those involving semi-leptonic PVES.
Results from the  SLAC deep-inelastic PVES experiment on deuterium were
reported in
the late 1970's \cite{Pre78}. These results also confirmed the structure of the
semi-leptonic weak neutral currents of the SM and yielded a value for the weak
mixing angle with nine uncertainty. About a decade later, the collaboration
at Mainz
reported results on a quasi-elastic PVES experiment involving a $^{8}$Be target
\cite{Hei89}. This experiment tested a different combination of the neutral
current
parameters than tested by the SLAC experiment. Shortly after the appearance
of the
Mainz result, the results of the elastic PVES experiment on $^{12}$C
performed at
Bates were repoted \cite{Sou90}. Again, the results of the carbon experiment
complemented those from quasi-elastic and deep inelastic measurements and
confirmed
the predictions of the SM. As discussed in more detail by Paul Souder, an
important
benefit of these PVES experiments was the development of experimental
expertise and
technology that is crucial to the sucess of the present program and the future
prospects of PVES.

Turning back to APV, the Boulder group's result for cesium dominates the
present
landscape. The group reports an experimental error of less than 0.4 \%. As
with the
earlier APV and PVES experiments, the goal of the cesium measurement was to
test
the SM. The cesium results deviate from the SM prediction by about
1.5\%, representing a 2.5$\sigma$ difference. The potentially serious
consequences
of this deviation call for a repeat measurement. To that end, the Bouchiat
group in
Paris is currently involved in another Stark-mixing experiment with cesium,
although the experimental uncertainty is not projected to be as small as in the
Boulder measurement.

Over the last decade, the emphasis of PVES has shifted away from SM tests
to the
study of hadron structure. As Paul Souder and Besty Beise discussed,  a
well-defined program of measurements to determine the nucleon's strange quark
vector current form factors is underway \cite{PVES89}. Instead of studying the
structure of the lepton-quark weak neutral current interaction, these
experiments
rely on the present knowledge of that interaction in order to learn
something new
about the sea-quark structure of the nucleon.  Results from the MIT-Bates
backward
angle experiments on the proton and deuterium have been reported by the SAMPLE
collaboration \cite{Mue97}, and the results of a forward angle measurement
have been
published by the HAPPEX collaboration at the Jefferson Lab \cite{Ani98}.
The list of
approved strange-quark experiments also includes the G0 experiment at the
Jefferson, a Hall-C experiment on $^4$He, and an experiment on the proton
at the
MAMI facility in Mainz \cite{PVES89}. The HAPPEX collaboration has also been
approved to run another forward angle proton measurement at $Q^2$ similar
to that
of the SAMPLE experiment. In addition, the G0 detector will be used to
measure the
axial vector $N\to\Delta$ transition form factor.

For both PVES and APV, the next generation of experiments are on the
horizon. The
groups in Seattle and Berkeley have undertaken measurements of APV
observables for
several atoms along the chain of isotopes. As I discuss below, {\em ratios}
of such
observables are less sensitive to atomic theory uncertainties than is the APV
observable for a single isotope. One hopes that such measurements may
provide an
even more precise tool for uncovering new physics than the Boulder cesium
experiment. In order to realize this goal, however, one requires a new level of
insight into {\em nuclear} structure than required for the interpretation of a
single isotope APV measurement. In the case of PVES, the interest of future
experiments seems to be returning to studying the weak neutral current
interaction
at the elementary fermion level. To that end, a purely leptonic experiment
involving PV M\"oller scattering has been approved for SLAC \cite{SLAC97}.
Similarly, a letter of intent to perform a precise, forward angle PV ${\vec
e}p$
experiment at the Jefferson Lab has appeared \cite{LOI99}. Finally, the
Jefferson
Lab PAC is considering a proposal to carry out elastic PVES with a $^{208}$Pb
target \cite{Mic99}. This experiment would provide the most precise
information we
have to date on the distribution of neutrons in a nucleus, something of
considerable interest to nuclear structure physicists. At the same time, 
the $^{208}$Pb experiment may provide enough nuclear structure information to help with the
interpretation of the APV isotope ratio studies in terms of new electroweak
physics. In this respect, the lead experiment would solidify a unique
marriage of table top and collider efforts having important consequences for atomic,
nuclear, and particle physics.

In the remainder of this discussion, I consider these future APV and PVES
experiments in detail. First, I review the motiviation for searching for new
physics at low-energies. I subsequently review the basics of the relevant PV
observables and show how precise measurements of these observables can
provide a
window on physics at the TeV scale. I give a few examples of new physics
scenarios
that can be tested by low-energy PV and consider a possible connection with
nuclear
$\beta$-decay. Finally, I discuss the relationship between the APV isotope
ratio
studies, the nuclear neutron distribution $\rho_n(r)$, and the PVES
experiment on
$^{208}$Pb. For an in-depth discussion of these issues, I refer the reader to
Refs. \cite{MRM99,Mus94}

\section{Searching for new physics}
\label{sec:newphys}
Although there exist a plethora of data confirming the electroweak sector
of the
Standard Model at the few $\times 0.1\%$ level, there also exist strong
conceptual
reasons to believe that the SM is only a piece of some larger framework.  A
nice
perspective from which to view the reasons for this belief is the so-called
high-energy desert. The high-energy desert is the region in mass scale
ranging from
the weak scale  $M_\sst{WEAK}\sim 250$ GeV up to the Planck scale
$M_\sst{P}\sim 1/\sqrt{8\pi G_\sst{NEWTON}}= 2.4\times 10^{18}$ GeV. The
conceptual
shortcomings of the SM appear at both edges of this desert. First, at the
high-energy end, the SM does not appear to produce unification of the
electroweak
and strong interactions at any scale. If one perturbatively runs the SU(3$)_C$,
SU(2$)_L$, and U(1$)_Y$ couplings up from the weak scale, they never meet at a
common point. This lack of unification is undesireable, particularly if one
believes a common framework ought to describe the electroweak, strong, and
gravitational interactions.

At the low-energy ($\mu << M_\sst{WEAK}$) edge ofthe desert, the SM is
similarly
less than satisfying. The most obvious shortcoming is the presence of 19
independent
parameters (in the limit of zero neutrino mass) which must be determined from
experiment. In addition, the violation of discrete symmetries, such as
parity and
CP, is put in by hand. The SM does not explain why nature violates these
symmetries; it simply incorporates them into a unified framework.
Similarly, the
quantization of electric charge must be put in by hand; it does not follow
naturally (at tree-level in the theory) as does, say, the quantization of
isospin
charge \cite{Moh92}. A particularly serious challenge for the SM is to
account for
the wide range of mass scales in the SM spectrum. A related aspect of this
\lq\lq
hierarchy problem" has to do with quadratic divergences appearing in the
renormalization of the Higgs mass. The presence of these divergences lead
one to
wonder why the Higgs mass should turn out to be at or below the weak scale
without the aid of some fine tuning of electroweak parameters. In short, the SM
leaves open many questions regarding the various mass scales governing
low-energy
physics.

Despite the phenomenological successes of the SM, then, one has good reason to
believe there must exist some larger framework which contains the
SU(3$)_C\times$SU(2$)_L\times$U(1$)_Y$ theory and which, presmumably, provides
answers to the conceptual puzzles of the SM. Hence, there exists intense
interest
these days in the search for new physics. In considering what such new physics
might be, one faces two broad questions: (a) Which new physics scenarios
are most
viable, both conceptually and phenomenologically? (b) What are the mass scales
associated with a given scenario? In the remainder of this discussion, I will
illustrate the insight parity-violating processes involving electrons might
play.

\section{PV observables}
\label{sec:pvobs}

The basic quantity of interest in considering neutral current PV is the
so-called
weak charge, $\qpv$. This quantity is the weak neutral current analog of the EM
charge. It gives the strength of the vector current coupling of $Z^0$-boson
to an
elementary fermion or system of fermions. For our purposes, it is useful to
write
the weak charge as
\begin{equation}
\label{eq:qpv}
\qpv = \qpv({\hbox{SM}}) + \Delta\qpv({\hbox{NEW}}) +
\Delta\qpv({\hbox{MB}})\ \ \ ,
\end{equation}
where the first term, $\qpv({\hbox{SM}})$ is the contribution to the weak
charge
from the SM. This contribution can be computed precisely and compared with an
experimental value for $\qpv$. Any significant deviation would signal non-zero
values for the remaining terms. Of these, $\Delta\qpv({\hbox{NEW}})$ represents
contributions from possible physics beyond the SM, while
$\Delta\qpv({\hbox{MB}})$
denotes contributions from convnetional many-body effects, such as strong
interactions among quarks which interfere with the $Z^0$-quark interaction. The
extent to which we can reliably compute the latter determines the
confidence with
which we can learn about $\Delta\qpv({\hbox{NEW}})$ from a given measurement.

Presently, the most precise determination of $\qpv$ has been obtained with
APV in
cesium. In an APV process, the weak neutral current interaction between the
electron and nucleus generates a PV atomic Hamiltonian which mixes states of
opposite parity in the atom:
\begin{equation}
\label{eq:hatom}
{\cal H}_\sst{W}^\sst{PV} = {\cal H}_\sst{W}^\sst{PV}({\hbox{NSID}}) +{\cal
H}_\sst{W}^\sst{PV}({\hbox{NSD}})\ \ \ .
\end{equation}
Here, \lq\lq NSID" and \lq\lq NSD" denote, respectively, nuclear spin-independent
and nuclear spin-dependent components of the 
interaction. The for
mer arises from
the product of axial vector electron and vector nuclear currents, whereas the
latter arises from a $V(e)\times A({\hbox{nucleus}})$ structure. These
terms can be
separated by measuring PV transitions between different hyperfine levels.
The NSID
term contains $\qpv$. The physics of the NSD term, which includes the
effects of
the nuclear anapole moment, is also interesting, though I will not consider it
further here (for a general discussion, see Ref. \cite{Mus91}).

As pointed out by the Bouchiats, the small parity-mixing effects caused by
${\cal
H}_\sst{W}^\sst{PV}$ can be enhanced by applying an electric field, which also
causes states of opposite parity to mix. Reversing the direction of the applied
field can isolate terms in the transition rate which depend on the
interference of
the Stark and weak interaction amplitudes. In the end, one extracts a ratio
such as
\begin{equation}
|A_\sst{PV}|/ |A_\sst{STARK}| = \xi\qpv\ \ \ ,
\end{equation}
where $\xi$ is an atomic structure-dependent constant that must be computed by
atomic theorists. Thus, any errors associated with atomic theory will propagate
into uncertainties in $\qpv$ (we might associate these conventional, atomic
structure uncertainties with $\Delta\qpv({\hbox{MB}})$ ). In fact, the dominant
uncertainty in the present value for $\qpv$ of cesium is from atomic theory.

An alternate method for determining $\qpv$ is with PVES. In PVES, one scatters
longitudinally polarized electrons from a target and compares the cross
sections
when the electron helicity is flipped. Any non-zero difference results from an
interference of the PV neutral current electron-nucleus scattering
amplitude and
the more familiar, parity-conserving electromagnetic amplitude. The
observable of interest in this case is the \lq\lq left-right" asymmetry
\begin{equation}
\label{eq:alr}
\alr = {N_{+} - N_{-}\over N_{+} + N_{-}} = a_0 Q^2\left\{{\qpv\over
Q_\sst{EM}} +
F(q)\right\} \ \ \ .
\end{equation}
Here, $N_{\pm}$ denote the number of electrons detected for a given
helicity of the
incident beam; $a_0$ is a constant whose scale is set by the Fermi and EM fine
structure constants; $Q^2$ is the square of the momentum transfer; $Q_\sst{EM}$
is the electromagnetic charge of the target; and
$F(q)$ is a term which depends on hadronic or nuclear form factors. In
principle,
one can separate the effects of $F(q)$ from those of $\qpv$ by exploiting the
kinematic dependence of the former. The goal of the present strange-quark
program
is to determined the contribution made by strange quarks to $F(q)$.

It is interesting to compare present and prospective determinations of
$\qpv$ with
those of other low-energy electroweak observables. In Table I I list several of
interest.

\begin{table}[h]
\caption{\quad Present and prospective limits on
low-energy electroweak observables. First three lines give present weak charge
limits for cesium and prospective precision for the SLAC M\"oller experiment
and possible Jefferson Lab experiment. Fourth line gives prospective isotope
ratio limits for APV studies in Seattle and Berkeley. Fifth line gives present
results for $|V_{ud}|^2$ for nine superallowed $\beta$-decays. Following
line gives
present and prospecive limits on the muon anomalous magnetic moment. Final
three lines give upper bounds on the permanent EDM's of the electron,
neutron, and
neutral mercury atom. Experimental errors are denoted by $(E)$ and theoretical
uncertainties by $(T)$.}
\begin{tabular}{cccc}
Observable & Quantity & Present Value & Source\\
\tableline
Weak Charge & $(Q_\sst{W}^\sst{EX}-Q_\sst{W}^\sst{SM})/Q_\sst{W}^\sst{WM}$
& $-0.016\pm0.0038 (E) \pm 0.005 (T)$& Cesium APV\\
 & &$ \hbox{?} \pm 0.07 (E) \pm 0.03 (T)$ & \hbox{PV ${\vec e}e$}\\
 & &$ \hbox{?} \pm 0.03 (E) \pm 0.03 (T)$ & \hbox{PV ${\vec e}p$} \\
 \hbox{Isotope Ratios} & $({\cal R}_\sst{EX} - {\cal R}_\sst{SM})/{\cal
R}_\sst{SM}$ & $\hbox{ ?} \pm 0.001 (E) \pm 0.004 (T)$ & \hbox{APV on Ba,Yb} \\
 \hbox{CKM Matrix} & $|V_{ud}|^2_\sst{EX}-|V_{ud}|^2_{SM}$& $-0.0028\pm 0.0013$& 
 \hbox{ $0^+\to 0^+$ $\beta$-decay} \\
 \hbox{Muon M.M.} & $\kappa_\mu^\sst{EX}-\kappa_\mu^\sst{SM}$ &$(750\pm 733)\times 10^{-11}$& \hbox{Present} \\
 & & $(\hbox{?}\pm 250)\times 10^{-11}$ & \hbox{BNL E821} \\
 \hbox{EDM} & $|d|$ & $\leq 4\times 10^{-28}\ e-\hbox{cm}$ &\hbox{electron}\\
 & & $\leq 0.97 \times 10^{-25}\ e-\hbox{cm}$ & \hbox{neutron} \\
& & $\leq 9\times 10^{-28}\ e-\hbox{cm}$ &$ ^{199}\hbox{Hg}$ \\
\end{tabular}
\end{table}

The top line in Table I gives the present limits on the agreement of the cesium
weak charge with the SM predicition. The Boulder group finds 2.5$\sigma$
deviation (about 1.5\%) from the SM value. The following rows give the expected
precision on the weak charge of the electron expected in the SLAC M\"oller
experiment and the weak charge of the proton in a prospective Jefferson Lab
experiment. It is worth noting that the electron and proton weak charges are
suppressed at tree level by $(1-4\sstw)\approx 0.1$; the electron weak
charge is
further suppressed by SM radiative corrections \cite{Cza96}. Crudely
speaking, then,
a 10\% determination of the proton or electron weak charge is equivalent to
a 1\%
determination of the weak charge of the cesium atom. The fourth line gives the
expected precision for the isotope ratio measurements at Berkeley and
Seattle. Note
that the prospective experimental error is much smaller than the present
theoretical uncertainty -- a point I address at the end of this discussion.

The other entries in Table I include the anomalous magnetic moment of the muon,
the permanent electric dipole moments (EDM's) of the electron, neutron, and
mercury atom; and the square of the $u-d$ matrix element of the CKM matrix.
Thus
far, one has no evidence of a non-vanishing permanent EDM or of a muon
anomalous
moment which differs from the SM prediction. In the case of $|V_{ud}|^2$,
however, an average over the results of nine superallowed, Fermi
nuclear $\beta$-decays yields a deviation from the requirements of CKM
unitarity
at the 2.2$\sigma$ level ( about 0.3\%) \cite{Tow95,Hag96}. It is
intriguing that
both semi-leptonic observables -- $\qpv$ from cesium APV and $|V_{ud}|^2$ from
superallowed
$\beta$-decay -- have the same relative sign for the experimental deviation
from
the SM prediction. If this discrepancy is due to new physics, this common
sign may
point to a common new physics scenario, as I discuss below.

\section{PV and new physics}
\label{sec:pvnewphys}

Before considering specific scenarios for physics beyond the SM, it is
useful to
consider the generic sensitivity of PV observable to such scenarios. In
doing so, I
follow the discussion of Ref. \cite{MRM99} restrict my attention to those
scenarios
which generate new effective, four-fermion interactions. Specifically, I write
the PV fermion-fermion interaction as
\begin{equation}
\label{eq:lfourfermi}
{\cal L}={\cal L}^\sst{PV}_\sst{S.M.}+{\cal L}^\sst{PV}_\sst{NEW}\ \ \ ,
\end{equation}
where
\begin{equation}
\label{eq:lstandard}
{\cal L}^\sst{PV}_\sst{S.M.}={G_F\over 2\sqrt{2}}\gae{\bar e}\gamma_\mu
	\gamma_5 e\sum_f \gvf{\bar f}\gamma^\mu f
\end{equation}
gives the SM contribution and
\begin{equation}
\label{eq:lpvnew}
{\cal L}^\sst{PV}_\sst{NEW}={4\pi\kappa^2\over\Lambda^2} {\bar e}
	\gamma_\mu\gamma_5 e\sum_f\hvf{\bar f}
	\gamma^\mu f
\end{equation}
is the contribution from some new physics. Here, $\gae$ axial vector
electron-$Z^0$
coupling and $\gvf$ is the vector current coupling of the $Z^0$ to fermion
$f$. In
Eq. (\ref{eq:lpvnew}), $\Lambda$ denotes the mass scale associated with the new
physics and $\kappa^2$ parameterizes the overall strength of the interaction. The
$\hvf$ give the scenario-specific couplings of 
the electron axial vector cu
rrent to
the vector current of fermion $f$. If the SM interaction in Eq.
(\ref{eq:lstandard}) determines the SM value of $\qpv$, the the fractional
shift
induced by the new interaction in Eq. (\ref{eq:lpvnew}) is
\begin{equation}
\label{eq:qpvshift}
{\Delta\qpv\over\qpv({\hbox{SM}})}={8\sqrt{2}\pi\over\Lambda^2 G_F}\ \ \ ,
\end{equation}
assuming $\gae\gvf$ and $\hvf$ have commensurate magnitudes.
If an experiment is sensitive to shifts on the order of
$\Delta\qpv/\qpv({\hbox{SM}})\sim 0.01$, then Eq. (\ref{eq:qpvshift})
implies one
is probing new physics at the $\Lambda\sim 20\kappa$ TeV scale. For new
physics of
a strong-interaction character, one expects $\kappa^2\sim 1$, while for new
gauge
interactions one expects $\kappa^2\sim \alpha$. In either case,
high-precision PV
measurements are incredibly powerful probes of physics at the TeV scale.

It is instructive to consider how these general features apply in the case of
specific new physics scenarios. One of the most interesting such scenarios
is that
of extended gauge symmetry. The basic of extended gauge symmetry is that the SM
group structure is embedded in some larger group $G$. The full symmetry of
$G$ may
break down spontaneously at one or more scales $M_\sst{X}$ above the weak
scale,
leaving the SU(3$)_C\times$SU(2$)_L\times$U(1$)_Y$ symmetry of the SM intact at
$M_\sst{WEAK}$. In principle, the gauge bosons associated with the additional
symmetries of $G$ will acquire masses commensurate with the symmetry
breaking scales
$M_\sst{X}$. If one of these scales is not too much larger than
$M_\sst{WEAK}$, then
the additional massive gauge bosons could generate small effects in low-energy
processes.

In addition to its phenomenological implications, extended gauge symmetry can
provide resolution to some of the rough edges of the SM. For example, if $G$
contains an SU(2$)_R$ subgroup, then one has a natural explanation for PV at
low-energies. At some high scale, one has exact parity symmetry. However,
if the
scale of symmetry breaking associated with the right-handed sector is much
larger
than $M_\sst{WEAK}$, the right-handed gauge bosons will be too heavy to compete
effectively with the SM gauge bosons, so that low-energy processes favor the
left-handed sector. Similarly, the electromagnetic charge can appear as a
generator
of $G$, in which case its quantization is natural. Even the apparent lack of
SM coupling unification can be resolved by extended gauge symmetry. The
presence of
additional symmetry breaking scales implies that the running of the
couplings will
change as one crosses each scale. Thus, there exists sufficient room within
different extended gauge group scenarios to bring about coupling
unification near
the expected grand unified scale.

Here, I concentrate on the neutral current phenomenology of extended gauge
symmetry. Specifically, I consider a scenario in which spontaneous symmetry
breaking of $G$ yields a second neutral gauge boson $Z'$  with mass not too
different from the weak scale. To make life simple,  I also consider the
case in
which this $Z'$ does not mix with the SM $Z^0$. If it did mix, its effects
would
show up strongly in the $Z$-pole observables. In fact, the latter severely
constrain the mass of a $Z'$ that does mix with the $Z^0$ \cite{Lan95}. In the
language of Eq. (\ref{eq:lpvnew}), we have for this scenario
$\kappa^2=\alpha'$, the
fine-structure constant associated with the $Z'$ interaction;
$\Lambda=M_\sst{Z'}$;
and the $\hvf$ to be specified by a particular scenario.

Given the experimental precisions listed in Table I, how sensitive would the
different measurements be to extended gauge symmetry-induced new interactions?
A detailed summary is given in Ref. \cite{MRM99}. Here, I quote a few
illustrative
examples. Extended gauge symmetry scenarios which fit naturally into the
framework
of heterotic strings live in a group called E$_6$. The factors of E$_6$
include two
U(1) groups called U(1$)_\chi$ and U(1$)_\psi$. The neutral gauge boson associated
with the U(1$)_\chi$ would show up 
particularly strongly in low-energy PV if
it had
a sufficiently low mass; the $Z_\psi$, on the other hand, does not
contribute to PV
amplitudes at tree-level. Let $G_\chi$ denotes the Fermi constant
associated with
the interactions of the $Z_\chi$. We may characterize the sensitivity of
various PV
observable in terms of the ratio $r_\chi = G_\chi/G_F$. The present cesium
APV is
able to discern effects of the scale $r_\chi \sim 0.003$ or larger. The
sensitivities of the SLAC M\"oller experiment, the proposed Jefferson Lab
PV $ep$
experiment, and the isotope ratio measurements are comparable. We can turn this
statement about Fermi constants into mass limits by assuming the break down of
E$_6$ to the SM  $\times$ U(1$)_\chi$ occurs in one step, so that the coupling
associated with the new U(1) group is maximal. In this case, the cesium APV,
isotope ratio, and PVES measurements would probe $M_{Z_\chi}$ at about the one
TeV level or better. In contrast, the sensitivity of the cesium measurement
to a
neutral right-handed gauge boson vastly exceeds the corresponding sensitivities
of the isotope ratio and PVES measurements. Thus, the use of different
low-energy PV measurements could prove useful in sorting out among competing
scenarios.

It is also interesting to compare the sensitivities of low-energy PV and
high-energy collider experiments. In terms of mass limits, the \lq\lq
reach" of the
present and prospective PV experiments exceeds that of the Tevatron by almost a
factor of two. Even an up-graded Tevatron (Tev33) would only achieve comparable
sensitivities. One must wait until the LHC has taken sufficient data before
the PV
sensitivities will be surpassed. In fact, the information provided by
colliders and
the PV measurements is complementary. The colliders are primarily sensitive
to the
mass scale associated with the new gauge boson relatively insensitive to the
coupling strength $g'$ or detailed structure of the fermion-$Z'$ coupling.
The PV
observables, in contrast, depend on $(g'/M_{Z'})^2$ ($\kappa/\Lambda$ in the
language of Eq. (\ref{eq:lpvnew})) and on the effective couplings fermion-$Z'$
couplings ($\hvf$ in Eq. (\ref{eq:lpvnew})).

To illustrate, I again consider E$_6$ theories \cite{Lon86}. The
phenomenology of
neutral E$_6$ gauge bosons is essentially governed by three parameters:
$M_{Z'}$; a
parameter
$\lambda_g$ which governs the overall coupling strength $g'$ and whose value
depends on the number of symmetry breaking steps leading to a massive $Z'$;
and an
\lq\lq extended" weak mixing angle $\phi$ which describes the structure of the
additional \lq\lq low-energy" U(1) group. Specifically, if $Z_\chi$ and
$Z_\psi$
are the gauge bosons associated with the U(1$)_\chi$ and U(1$)_\psi$ groups,
respectively, then a general neutral E$_6$ gauge boson can be written as
\begin{equation}
\label{eq:znew}
Z'= \cos\phi Z_\psi + \sin\phi Z_\chi \ \ \  .
\end{equation}
The couplings $\hvf$ of this $Z'$ to electrons and light quarks are given by
\begin{eqnarray}
\label{eq:zlight}
\hvu&=&0 \\
\hvd=-\hve&=&\left[\sin^2\phi-\sqrt{15}\sin\phi\cos\phi/3\right]/20\ \ \ .
\end{eqnarray}
Note that for $\phi=0$ or $\pi$, $Z'=Z_\psi$ and all of the PV couplings
vanish.
The d-quark and electron couplings also vanish for
$\phi=\phi_c=\tan^{-1}(\sqrt{5/3})$ and have opposite signs for $\phi$ on
either
side of $\phi_c$. Thus, the net effect of the $Z'$ on $\qpv$ can be either
positive
or negative, depending on the value of $\phi$. The present present cesium APV
results favor $\phi>\phi_c$, if an E$_6$ gauge boson is responsible for the
observed
deviation from the SM value for $\qpv$. This kind of information about the
structure of the extended gauge sector is difficult to obtain from high-energy
collider limits.

It is also amusing to combine information obtained from colliders and
low-energy
experiments. To do so, let's assume the E$_6$ gauge boson is responsible
for the
deviation of the cesium $\qpv$ from the SM value (about a two $\sigma$
effect).
Under this assumption, one has a relationship between $M_{Z'}$,
$\lambda_g$, and
$\phi$. A second condition derives from the CDF lower bounds, which are roughly
600 GeV with little dependence on the value of $\phi$. Combining the two
pieces of
information, one obtains
\begin{equation}
\label{eq:zlimits}
600\ \ {\hbox{GeV}} \lsim M_{Z'} \lsim 1.15\lambda_g \ \  {\hbox{TeV}}\ \ \ ,
\end{equation}
where $\lambda_g\leq 1$. This range is already rather narrow. If a future
up-graded Tevatron found no evidence for extra neutral gauge bosons with a mass
less than about one TeV, then  a low-mass $Z'$ would be ruled out as the
culprit
behind the cesium APV result.

Another popular extension of the Standard Model is supersymmetry. The
literature on
SUSY extensions of the SM is legion, so I will not discuss SUSY models in
detail.
The appeal of SUSY includes its solution to the hierarchy problem
associated with
mass renormalization. In addition, the gauge couplings in the minimal
supersymmetric standard model (MSSM) unify at the GUT scale when run
perturbatively
up from the weak scale. Whether this coupling unification is fortuitous or
reflects
deeper physics can be debated. It is, nevertheless, intriguing. One important
characteristic of the MSSM as far as low-energy phenomenology is concerned
involves
a quantity called R-parity. The R-parity quantum number  is defined as
\begin{equation}
P_R = (-1)^{3(B-L)+2S}\ \ \ ,
\end{equation}
where $B$, $L$, and $S$ denote the baryon number, lepton number, and spin,
respectively, of a given particle. Every SM particle has $P_R=1$ while each
superpartner has $P_R=-1$. The MSSM conserves total $P_R$, which implies
that every
interaction involves an even number of superpartners. As a result,
superpartners
cannot appear in low-energy processes involving SM particles at tree-level.
They
only contribute through loops. Their effects are correspondgingly suppressed by
loop factors, making them hard to see at low-energies.

It is possible, however, to write down simple extensions of the MSSM in
which $P_R$
is not conserved. In such $B$ and/or $L$-violating theories, superpartner
effects
can appear at tree-level. To illustrate, consider a purely leptonic R
parity-violating SUSY model. The relevant Lagrangian is \cite{Bar89}
\begin{equation}
\label{eq:rparity}
{\cal L}_\sst{RPV} = \lambda_{ijk} ({\tilde e}^k_R)^{\ast}\ ({\bar\nu}^i_L)^c\
e_L^j \ + \ {\hbox{h.c.}}\ \ \ ,
\end{equation}
where ${\tilde e}^k_R$ denotes the bosonic superpartner of a right-handed
charged
lepton of generation $k$ (the other superscripts denote generation). Since the
interaction contains three leptons, $L$ (and $P_R$) are not conserved.
Tree-level
exchange of the ${\tilde e}^k_R$ between lepton currents can generate new
four-fermion effective interactions, such as the following interaction
relevant to
$\mu$-decay:
\begin{equation}
\label{eq:rpvb}
{\cal L}_\sst{EFF} = -(\lambda_{12k}/\sqrt{2} M_{\phi^e_{kR}})^2 {\bar
e}_L\gamma_\alpha
	\nu^e_L {\bar\nu}^\mu_L \gamma^\alpha \mu_L \ \ \ .
\end{equation}

The interaction of Eq. (\ref{eq:rparity}) may provide a partial explanation for
both the cesium APV result and the apparent CKM unitarity violation
inferred from
the superallowed $\beta$-decays. The reason has to do with the Fermi constant.
Both the $\beta$-decay amplitude and the PV amplitude of Eq.
(\ref{eq:lstandard})
are written in terms of the Fermi constant. The reason is that these amplitudes
depend on $g^2/\mws$, which can be related to the Fermi constant as measured in
$\mu$-decay. At tree-level, this relationship is given by
\begin{equation}
\label{eq:gftree}
{g^2\over 8\mws} = {G_F\over\sqrt{2}}\ \ \ .
\end{equation}
Because of the precision with which $\mu$-decay is measured, Eq.
(\ref{eq:gftree}) must be modified to account for electroweak radiative
corrections:
\begin{equation}
\label{eq:gfrad}
{g^2\over 8\mws}(1+\Delta r) = {G_\mu\over\sqrt{2}}\ \ \ ,
\end{equation}
where $\Delta r$ contains the radiative corrections. Suppose now some new
physics,
such as the interaction of Eq. (\ref{eq:rpvb}), contributes to $\mu$-decay.
Then
one must further modify Eq. (\ref{eq:gfrad}) as
\begin{equation}
\label{eq:gfnew}
{g^2\over 8\mws}(1+\Delta r+\dnewmu) = {G_\mu\over\sqrt{2}}\ \ \ ,
\end{equation}
where $\dnewmu$ gives the corrections from the new interaction. When
writing down
the amplitude for $\beta$-decay or PV, one needs $g^2/\mws$ in terms of
$G_\mu$:
\begin{equation}
\label{eq:gsquared}
{g^2\over 8\mws} = {G_\mu\over\sqrt{2}}(1-\Delta r-\dnewmu)
\end{equation}
to first order in the small corrections.

To make contact with the semi-leptonic observables, it is useful to
consider the
effective Fermi constants $G_F^\beta$ and $G_F^\sst{PV}$ which govern them. In
terms of other quantities, these effective Fermi constants are
\begin{eqnarray}
\label{eq:gfeffective}
G_F^\beta&=& G_\mu(1-\Delta r +\Delta r_\beta - \dnewmu+\dnewbeta) |V_{ud}|^2\\
G_F^\sst{PV} &=& G_\mu(1-\Delta r +\Delta r_\sst{PV} - \dnewmu+\dnewpv) \qpv
\ \ \ ,
\end{eqnarray}
where $\Delta r_\beta$ and $\Delta r_\sst{PV}$ denote SM radiative
corrections to
the $\beta$-decay and PV amplitudes, respectively, and $\dnewbeta$ and
$\dnewpv$
are the corresponding contributions from new interactions.

The results of from the superallowed decays and cesium APV imply
\begin{eqnarray}
G_F^{\beta , \sst{EX}}/G_F^{\beta ,\sst{SM}} &<& 1 \\
G_F^{\sst{PV} ,\sst{EX}}/G_F^{\sst{PV} ,\sst{SM}} &<& 1
\end{eqnarray}
where the $EX$ and $SM$ superscripts denote the experimental and SM values,
respectively. From Eq. (\ref{eq:gfeffective}), we see that if the new physics
contributions vanish, one obtains the conventional interpretation of the
experimental results:
\begin{eqnarray}
\label{eq:semiconv}
|V_{ud}|^2_\sst{EX}/|V_{ud}|^2_\sst{SM} &<& 1 \\
Q_\sst{W}^\sst{EX}/Q_W^\sst{SM} &<& 1 \ \ \ .
\end{eqnarray}
However, an equally acceptable explanation is to assume $|V_{ud}|^2$ and
$\qpv$
assume their SM values and that
\begin{eqnarray}
\dnewbeta-\dnewmu &<& 1 \\
\dnewpv -\dnewmu &<& 1\ \ \ .
\end{eqnarray}
In particular, if both $\dnewbeta$ and $\dnewpv$ vanish and if $\dnewmu>0$, the
measured effective Fermi constants in $\beta$-decay and cesium APV would be
smaller
in magnitude than the SM predictions.

The R parity-violating interaction of Eq. (\ref{eq:rpvb}) generates just such a
positive value for $\dnewmu$:
\begin{equation}
\label{eq:dnewmususy}
\dnewmu = {\lambda^2_{12k}\over 4\sqrt{2}G_\mu M^2_{\phi^e_{kR}}}\ \ \ .
\end{equation}
Using the present experimental results and Eq. (\ref{eq:gfeffective}) one
obtains
\begin{equation}
\label{eq:rpvbeta}
\lambda_{12k} = (0.027\pm 0.007) (M_{\tilde e_k}/100\ {\hbox{GeV}})
\end{equation}
from superallowed decays and
\begin{equation}
\label{eq:rpvparity}
\lambda_{12k} = (0.13\pm 0.05) (M_{\tilde e_k}/100\ {\hbox{GeV}})
\end{equation}
from cesium APV. Although these results differ by more than one $\sigma$, one
should keep in mind that the cesium result is the first PV result to differ
from the
SM, whereas the superallowed results depend on an average of $ft$ values
for nine
different decays, several of which have been measured more than once. In
short, the
precise magnitude of the deviation leading to Eq. (\ref{eq:rpvparity}) may
not be
as robust as that observed in $\beta$-decay. The primary point here is that the
magnitudes of the results in Eqs. (\ref{eq:rpvbeta}-\ref{eq:rpvparity}) are
not too
distinct, and the signs of the observed deviations are both consistent with the
R parity-violating effects in Eqs. (\ref{eq:rpvb}) and (\ref{eq:dnewmususy}).
It will be interesting to see whether future electron PV experiments also
produce
deviations from the SM predictions consistent with this SUSY
scenario\footnote{Another constraint on R parity-violating SUSY may be
obtained from
relations among electroweak parameters. The constraints imposed by these
relations
on some types of new physics have been analyzed in Ref. \cite{Mar99}. The
corresponding SUSY constraints will be discussed in a forthcoming publication}.

\section{interpretation issues and neutron distributions}
\label{sec:interp}

In general, the interpretation of precision, low-energy measurements raises
thorny
issues not relevant to high-energy measurements. The PV processes discussed
here
are no exception. To illustrate, I consider the interpretation of atomic PV. As
noted above, the dominant error in the cesium weak charge comes from atomic
theory.
Although this theory error appears to have been reduced in light of new
measurements of parity-conserving atomic transitions, it is questionable
whether
further reductions can be achieved. A clever strategy for evading this atomic
structure uncertainty is to measure ratios of APV observables along an isotope
chain. A representative ratio is
\begin{equation}
{\cal R} = {\apvnsid(N')-\apvnsid(N)\over\apvnsid(N')+\apvnsid(N)}\ \ \ ,
\end{equation}
where $\apvnsid(N)$ is an APV nuclear spin-independent observable for an
atom with
neutron number $N$. Since the atomic electronic structure contributions
$\apvnsid(N)$ and $\apvnsid(N')$ are relatively constant (for a given $Z$), the
atomic structure-dependence drops out of the ratio ${\cal R}$ and one has
\begin{equation}
{\cal R} \approx {\qpv(N') - \qpv(N)\over \qpv(N) + \qpv(N')}\equiv
{\cal R}\sst{SM} (1+\delta_{\cal R}) \ \ \ ,
\end{equation}
where ${\cal R}_\sst{SM}$ is the value of the ratio in the SM.

The correction $\delta_{\cal R}$ contains contributions from possible new
physics.
As first pointed out by Fortson, Wilets, and Pang, however, there is also a
second
effect due to the variation of the neutron density $\rho_n(r)$ along the
isotope
chain \cite{For90}. To get an idea of the relative importance of these two
contributions, one can model the nucleus as a sphere of constant neutron
and proton
density out to radii
$R_N$ and $R_P$, respectively. In this case, one has
\begin{equation}
\label{eq:ratio}
\delta_{\cal R} \approx \left({2Z\over N+N'}\right)\Delta\qwp -
\left({N'\over\Delta N}\right) (Z\alpha)^2 (3/7) \delta(\Delta X_N)\ \ \ ,
\end{equation}
where $\Delta\qwp$ is the shift in the proton's weak charge due to new physics,
\begin{equation}
\Delta X_N= {R_{N'} - R_N\over R_P}\ \ \
\end{equation}
is the shift in the mean square neutron radius (relative to the proton radius)
along the isotope chain,  and $\delta(\Delta X_N)$ is the uncertainty in this
shift.

Several features of Eq. (\ref{eq:ratio}) are worth noting. First, the shift in
the ratio ${\cal R}$ due to new physics depends primarily on the shift in
the weak
charge of the proton. The shift in the weak charge of the neutron largely
cancels
out of the ratio, to first order in small shifts. Whereas the weak charge of a
single isotope is slightly more sensitive $\Delta\qwn$ than to $\qwp$, the
sensitivity of
${\cal R}$ to new physics is dominated by $\Delta\qwp$. Second, the
dependence of
${\cal R}$ on variations in neutron radii along the isotope shift is
enhanced by
a factor of $N'/\Delta N$. For a heavy atom like cesium or barium, for example,
this enhancement factor can be on the order of 5. Thus, if one is going to use
APV isotope ratio measurements to learn about $\Delta\qwp$, one must have
extremely
precise knowledge of the shift in neutron radii.

At present, there exist no high-precision experimental determinations of
the neutron
radii of heavy nuclei. Consequently, nuclear theory must be used to
determine the
second term on the RHS of Eq. (\ref{eq:ratio}). To set the scale of the
level of
accuracy nuclear theory must achieve to make the isotope ratio measurements
useful,
supposed we require the uncertainty in the neutron radius term to be as
small as the
prospective experimental uncertainty in the value of ${\cal R}$, namely,
0.1 \%.
Pollock \cite{Pol92} and Chen and Vogel \cite{Che93,Vog94} have analyzed
the nuclear
model spread in $\Delta X_N$; from their analyses, we learn that nuclear
theory is
at least a factor of two away from achieivng the requisite precision (for a
summary
of the theoretical situation, see Ref. \cite{MRM99}). In principle, this
presents a
stumbling block for the isotope ratio program.

There exist two strategies for overcoming this difficutly. One is to perform a
direct measurement of $\Delta\qwp$ using PVES from a proton target. From Eq.
(\ref{eq:alr}), we may write the proton asymmetry as
\begin{equation}
\alr(^1{\hbox{H}}) = a_0 Q^2\left[\qwp+F^p(q)\right]\ \ \ ,
\end{equation}
where $\qwp$ is the proton weak charge. The form factor term $F^p(q)$ vanishes
in the forward angle limit. Thus, by going to forward angle kinematics, the
$\qwp$ can be separated from $F^p(q)$. The form factor term is presently under
study in the strange quark experiments. Upon completion of the strange quark
program, this term should be known with sufficient precision over a large
enough
kinematic range to afford a precise separation of $\qwp$ in a future,
forward angle
measurement. A letter of intent for such a measurement has recently been issued
\cite{LOI99}. The proposed measurement would employ a re-configured G0
apparatus in
order to reach suffienct forward angle kinematics. It is hoped that this
measurement will yield a 3-5\% determination of $\qwp$. This level of precision
would be comparable to a 0.1-0.2\% determinatio of ${\cal R}$, if the
interpretation of the latter were not clouded by $\rho_n(r)$ uncertainties.

A second for getting around the $\rho_n(r)$ problem in Eq. (\ref{eq:ratio})
involves measuring the neutron distribution of a heavy nucleus using PVES.
It is possible that a sufficiently precise determination of $\rho_n(r)$ on
a single isotope would sufficiently constrain nuclear theory that the nuclear
model-dependence in the isotope shifts, $\delta (\Delta X_N)$ would be reduced
to an acceptable level. The idea for using PVES to determine $\rho_n(r)$ was
first suggested by Donnelly, Dubach, and Sick \cite{Don89}. These authors noted
that the $Z^0$ preferentially sees neutrons over protons, since at tree-level
in the SM, $\qwp = 1-4\sstw\sim 0.1$ whereas $\qwn = -1$. Thus, the PV
asymmetry
for scattering from a heavy nucleus should be quite sensitive to the neutron
distribution. To illustrate this idea, consider PVES from a $(J^\pi,
T)=(0^+,0)$
nucleus. The asymmetry has the form \cite{Don89,Mus94}
\begin{equation}
\label{eq:heavyasym}
-\left[{4\sqrt{2}\pi\alpha\over G_F|Q^2|}\right]\alr = \qwp+\qwn{\int\ d^3x\
j_0(qx) \rho_n({\vec x}) \over \int\ d^3x\ j_0(qx) \rho_p({\vec x})}\ \ \ .
\end{equation}
Since $\rho_p({\vec x})$ is typically known with very high accuracy, the
PV asymmetry essentially becomes a \lq\lq meter" of $\rho_n(q)$. This idea
is being
exploited in a proposal before the Jefferson Lab PAC \cite{Mic99}.

It goes without saying that a precise determination of $\rho_n(q)$ for any
heavy
nucleus is of fundamental interest for nuclear structure physics. From this
standpoint alone, the investment of effort in making the measurement is
well-justified. It remains to be seen, however, whether the information
gleaned from
a precise determination of $\rho_n(q)$ for $^{208}$Pb at one or two kinematic
points will suffice to reduce the nuclear structure uncertainty in Eq.
(\ref{eq:ratio}). For example, it is unlikely that lead atoms will be used
in the
APV isotope ratios. The isotopes of Ba and Yb are currently under study in
Seattle
and Berkeley. Moreover, the interpretation of ${\cal R}$ requires knowledge of
$\rho_n(r)$ in more detail than implied by the simplified expression in Eq.
(\ref{eq:ratio}). Whether knowledge of the momentum-space
distribution at a few points will supply the necessary details about
$\rho_n(r)$
is an open question.  Finally, the constraints which knowledge of
$\rho_n(r)$ for a
single isotope would place on calculations of isotope shifts has yet to be
quantified. In short, there exist several challenges for nuclear theory in
making
a PVES determination of $\rho_n(q)$ useful for the APV isotope ratios (for
a recent
discussion of these issues, see Ref. \cite{Hor99}). From this standpoint, a
measurement of the PV ${\vec e}p$ asymmetry provides a cleaner and more direct
window on $\Delta\qwp$.

\section{Conclusions}
\label{sec:conclusions}

The field of parity-violation with electrons has made tremendous strides in 25
years. I hope this discussion has convinced the reader that its future
prospects
are just as exciting as its history. For the next decade at least, it is likely
that PV with electrons will provide one of the most powerful probes of new
physics
at the TeV scale, complementing information to be gained from high-energy
collider
experiments. At the same time, it will remain a focal point for
interdisciplinary
activity, bringing together insights from particle, nuclear, and atomic
physics.
One may only speculate as to the {\em new} insights PV with electrons will
provide
for each field by the time a Bates-35 celebration is planned.

\begin{center}
{\bf ACKNOWLEDGEMENTS}
\end{center}

It is a pleasure to thank W.J. Marciano, D. Budker, R. Carlini, J.M. Finn,
E.N. Fortson, S.J.
Pollock, and P. Souder for useful discussions and S.J. Puglia for
assistance in preparing the
manuscript. This work was supported in part under U.S. Department of Energy
contract
\#DE-AC05-84ER40150 and a National Science Foundation Young Investigator
Award.

\vfill
%\eject

\end{document}